\documentclass[aps,pra,amsmath,amssymb,amsfonts,lengthcheck,twocolumn,superscriptaddress]{revtex4-2}
\usepackage{graphicx}
\usepackage{subfigure}
\usepackage{amsmath, amsthm, amssymb, amsfonts}
\usepackage{verbatim}
\usepackage{dcolumn}
\usepackage{bm}
\usepackage{epsf}
\usepackage{color}
\usepackage[colorlinks=true,citecolor=blue,linkcolor=blue,urlcolor=blue]{hyperref}%
\usepackage{xcolor}
\usepackage{physics}
\usepackage{dsfont}
\usepackage{tikz}
\usepackage{multirow}
\usepackage{xfrac}
\usepackage{makecell}
\usepackage{algorithm}
\usepackage[noend]{algpseudocode}
\usepackage{siunitx}
\usepackage{orcidlink}
\usepackage{todonotes}

\newcommand{\mtr}[1]{\mathrm{tr}\left\{#1\right\}}

\newcommand{\bla}{bla\\bla\\bla\\bla\\bla}

\theoremstyle{definition}

\newtheorem{theorem}{Theorem}[section]
\newtheorem{corollary}{Corollary}[section]

\DeclareMathAlphabet\mathbfcal{OMS}{cmsy}{b}{n}

\makeatletter
\newcommand{\currentfontsize}{\f@size pt}
\makeatother

\makeatletter
\newcommand\footnoteref[1]{\protected@xdef\@thefnmark{\ref{#1}}\@footnotemark}
\makeatother

\begin{document}


\title{Clifford Ergotropy -- determining the energetic cost of generating quantum magic}

\author{Noah B. Kerzner\orcidlink{0009-0009-8155-266x}}
\affiliation{Department of Physics \&\ Astronomy, Bucknell University, Lewisburg, PA 17837, USA}

\author{Josey Stevens\orcidlink{0000-0003-0627-1592}}
\affiliation{Department of Physics, University of Maryland, Baltimore County, Baltimore, MD 21250, USA}
\affiliation{Quantum Science Institute, University of Maryland, Baltimore County, Baltimore, MD 21250, USA}

\author{Emery Doucet\orcidlink{0000-0002-2693-8553}}
\affiliation{Department of Physics, University of Massachusetts, Boston, Massachusetts 02125, USA}

\author{Benjamin Vollmayr-Lee\orcidlink{0000-0002-3794-1248}}
\affiliation{Department of Physics \&\ Astronomy, Bucknell University, Lewisburg, PA 17837, USA}

\author{Sebastian Deffner\orcidlink{0000-0003-0504-6932}}
\affiliation{Department of Physics, University of Maryland, Baltimore County, Baltimore, MD 21250, USA}
\affiliation{Quantum Science Institute, University of Maryland, Baltimore County, Baltimore, MD 21250, USA}
\affiliation{National Quantum Laboratory, College Park, MD 20740, USA}


\date{\today}

\begin{abstract}
The core resource required to achieve quantum advantage has been dubbed \emph{quantum magic}. However, quantum thermodynamics dictates that no informational resources are for free and that their creation and assumption has an corresponding energetic cost. For quantum magic this cost is determined by the Clifford ergotropy. However, the superexponential growth of the Clifford group as a function of the dimension of the logical space makes the exhaustive evaluation computationally prohibitive, while its conjugation constraints complicate analytical solutions. In this work, we rephrase the problem as a search over the Hamiltonian’s Clifford orbit and show how commutation structure and linear independence of its Pauli terms constrain this orbit. As a main result and using the binary symplectic formalism, we develop an approximation scheme with explicit error bounds for Hamiltonians whose Pauli terms are linearly independent over the binary field.

\end{abstract}

\maketitle


\section{Introduction}

Recent developments in quantum thermodynamics have illuminated the role that nonclassical resources can play in work extraction~\cite{Goold2016,Llobet2015PRX,Myers2021PRXQ,Touil2021Entropy,Touil2022JPA,Sone2021entropy,Aguilar2025SA}. In particular, quantum properties of the system's state, such as coherence and entanglement, can influence the amount of energy that is operationally accessible~\cite{Lostaglio2015,Francica2017,Alipour2016SR}. 

Defining work in quantum systems has been the subject of considerable discussion, cf. Ref.~\cite{Kalsi2026fqxi} for a recent review on the topic. However, for an isolated quantum system driven by an external, time-dependent control field, work can be identified with the change in the system's average energy. In the absence of interactions with an environment, the system evolves unitarily, and the work extracted is given by the decrease in its average energy. When the external driving is cyclic, such that the Hamiltonian returns to its initial form, the maximum work extractable through unitary transformations is quantified by the ergotropy~\cite{Allahverdyan2004},
\begin{equation}
    \mathcal{E}(\rho,H) =\mtr{\rho H}-\min_{U\in\mathcal{U}(d)}\mtr{U\rho U^\dagger H}\,,
\end{equation}
where the minimization is taken over all unitary operations on the system.

Under the assumption of unrestricted cyclic unitary control, ergotropy completely characterizes the work-extraction potential of a quantum state~\cite{Allahverdyan2004}. In practice, however, arbitrary unitary transformations may be inaccessible or prohibitively costly, particularly in large many-body systems, where realizing generic global unitaries requires increasingly complex and precise control over many interacting degrees of freedom. Indeed, the experimental complexity of implementing the optimal global unitary has motivated the consideration of more restricted and physically accessible control protocols~\cite{Brown2016,Salvia2023,Castellano2024}. Previous works have considered restrictions arising from locality and limited subsystem control~\cite{Salvia2023,Castellano2024}, as well as, in continuous-variable systems, restriction to Gaussian unitaries~\cite{Brown2016}, with each restriction defining an operationally constrained notion of extractable work.

A particularly natural restricted set arises in gate-based quantum computing, where the various unitary operations can have markedly different computational costs. Clifford operations generate stabilizer circuits that can be efficiently simulated classically~\cite{Aaronson2004}, whereas non-Clifford resources provide the magic (i.e., non-stabilizerness) required to promote otherwise Clifford circuits to universal quantum computation~\cite{Bravyi2005}. In fault-tolerant architectures, the implementation of non-Clifford operations can moreover constitute a substantial resource overhead, motivating extensive study of their computational cost~\cite{Bravyi2005,Beverland2021}. The thermodynamic consequences of non-Clifford operations, however, remain comparatively less well understood. It is therefore natural to ask how restricting the available control operations to the Clifford group affects the amount of extractable work.

Reference~\cite{maity2026clifford} recently formalized this question by introducing the Clifford ergotropy, which, for $H$ being the system Hamiltonian, $\text{Cl}_n$ is the $n$-qubit Clifford group, and $\rho$ is the system state. It is defined by
\begin{equation}
    \mathcal E_{\mathrm{Cl}}(\rho,H)
    =
    \mtr{\rho H}
    -
    \min_{U\in\mathrm{Cl}_n}
    \mtr{U\rho U^\dagger H}\,.
    \label{eq:clifford-ergotropy}
\end{equation}
Note that $\mathcal E_{\mathrm{Cl}}(\rho,H)$ naturally quantifies the energetic cost associated with generating quantum magic, and hence quantum advantage. However, actually evaluating $ \mathcal E_{\mathrm{Cl}}(\rho,H)$ is a computationally involved problem, since it is equivalent to quantifying the present quantum advantages, which generally requires a quantum computer.

Therefore, Ref.~\cite{maity2026clifford} focused on a simple and tractable situation. Namely, the authors derived  an analytic expression for a single qubit under a one-term Hamiltonian, compared Clifford ergotropy with several established magic monotones, and investigated its behavior in a number of small many-body systems. Despite the simplicity of the considered scenarios, Ref.~\cite{maity2026clifford}  established Clifford ergotropy as a useful thermodynamic probe of non-Clifford resources.

In our present work, we extend the formalism introduced by Ref.~\cite{maity2026clifford} by finding classes of Hamiltonians for which evaluating the Clifford ergotropy becomes computationally efficient. In fact, we will see that special commuting or anti-commuting structures of the underlying Hamiltonian allows to determine $\mathcal E_{\mathrm{Cl}}(\rho,H)$ in closed form.

This paper is organized as follows: In Sec.~\ref{sec:definitions}, we introduce the Clifford ergotropy and give several important definitions and simplifications. Then we derive an analytic expression for the one-qubit ergotropy in Sec.~\ref{sec:one_qubit-ergo},  and demonstrate the fundamental challenge in analytically finding the quantity. In Sec.~\ref{sec:one_commutation_relation}, we impose a single commutation restraint on the input Hamiltonian and show that finding the Clifford ergotropy becomes significantly easier. Next,  in Sec.~\ref{sec:binary_symplectic}, we introduce the binary symplectic formalism to identify a broad class of Hamiltonians which admit a simplified calculation. Then we analyze features of the solvable classes of Hamiltonians  in Sec.~\ref{sec:evaluating_ergo} and offer an approximation method for the Clifford ergotropy. Finally, in Sec.~\ref{sec:conclusion} we conclude this work and point to future extensions. 

\section{The Clifford Group and the Clifford Ergotropy}
\label{sec:definitions}

We start by establishing notions and notations. To this end, let $\mathcal P_n$ denote the $n$-qubit Pauli group,
\begin{equation}
    \mathcal P_n
    =
    \left\{
        i^k \sigma_1\otimes\cdots\otimes \sigma_n|\,
        k\in\{0,1,2,3\}\right\}.
\end{equation}
where $\sigma_j\in\{I,X,Y,Z\}$ are the Pauli matrices. The $n$-qubit Clifford group is the normalizer of the Pauli group
\begin{equation}
    \mathrm{Cl}_n=
    \left\{U:U^\dagger \sigma_i U\in\mathcal P_n,  \forall \sigma_i\in\mathcal P_n \right\}.
\end{equation}
Clifford conjugation maps every Hermitian Pauli string to another Hermitian Pauli string up to sign,
\begin{equation}
    U^\dagger \sigma_i U=s_i \sigma'_i,
    \quad\text{with}\quad
    s_i\in\{\pm1\},
\end{equation}
where $\sigma'_i\in\mathcal P_n^{\mathrm H}$. The Clifford ergotropy, introduced in Ref.~\cite{maity2026clifford}, is then defined via Eq.~(\ref{eq:clifford-ergotropy}). 
For convenience, we define the Clifford minimum energy
\begin{equation}
    E_{\mathrm{Cl}}^{\min}(\rho,H)
    :=
    \min_{U\in\mathrm{Cl}_n}
    \mtr{U\rho U^\dagger H},
\end{equation}
and hence that the Clifford ergotropy can be written as
\begin{equation}
    \mathcal E_{\mathrm{Cl}}(\rho,H)
    =
    \mtr{\rho H}-E_{\mathrm{Cl}}^{\min}(\rho,H).
\end{equation}
By cyclicity of the trace, $ E_{\mathrm{Cl}}^{\min}(\rho,H)=\min_{U\in\mathrm{Cl}_n}\mtr{\rho U^\dagger H U}, $ which motivates the introduction of the Clifford orbit of the Hamiltonian \cite{maity2026clifford},
\begin{equation}
    \mathcal O_{\mathrm{Cl}}(H):=\left\{U^\dagger H U: U\in\mathrm{Cl}_n\right\}.
\end{equation}
The Clifford minimum may then be written as
\begin{equation}
    E_{\mathrm{Cl}}^{\min}(\rho,H)
    =
    \min_{H'\in\mathcal O_{\mathrm{Cl}}(H)}
    \mtr{\rho H'}\,.
    \label{eq:hamiltonian_clifford_minimum}
\end{equation}
Therefore, evaluating the Clifford ergotropy is equivalent to enumerating the energy over every Hamiltonian in $\mathcal O_{\mathrm{Cl}}(H)$ and selecting the minimum. However, in practice, such a task is exponentially hard. The order of the $n$-qubit Clifford group, modulo global phases, is given in Ref.~\cite{Koenig2014} by
\begin{equation}
|\mathrm{Cl}_n|
=2^{n^2+2n}\prod_{j=1}^{n}\left(4^j-1\right)
=\mathcal{O}\left(2^{2n^2+3n}\right),
\label{eq:clifford_group_size}
\end{equation}
which is super-exponential in $n$. Enumerating through the group for even modestly sized $n$ is then computationally intractable. 

In the following, we, thus, investigate classes of Hamiltonians with Clifford orbits that grow more slowly than the full Clifford group. In particular, we consider specific algebraic conditions for certain types of Hamiltonians under which the orbit is sufficiently small to make the direct calculation of Clifford ergotropy computationally feasible. 

Furthermore, consider that the Hermitian Pauli strings form an orthogonal basis for the space of
$n$-qubit Hermitian operators. Any Hamiltonian and density
operator may be expanded as
\begin{equation}
    H=\sum_{\sigma_i\in\mathcal P_n^{\mathrm H}}a_i \sigma_i,
    \quad \text{and}\quad
    \rho=\frac{1}{2^n}
    \sum_{\sigma_i\in\mathcal P_n^{\mathrm H}}r_i \sigma_i,
    \label{eq:pauli_expansions}
\end{equation}
where $\mathcal P_n^{\mathrm H}=\{I,X,Y,Z\}^{\otimes n}$ denotes the unsigned Hermitian Pauli basis, $a_i\in\mathbb R$, and $r_i=\mtr{\rho \sigma_i}$. We write $\mathcal P_n^*=\mathcal P_n^{\mathrm H}\setminus\{I\}$. Throughout, $a$ denotes Hamiltonian coefficients and $r$ denotes state Pauli coefficients; for one qubit, $\vec r=(r_X,r_Y,r_Z)$ is the Bloch vector. The symbols $r_{(i)}(\mathcal Q)$ denote sorted magnitudes within a specified unordered set $\mathcal Q$, while $R_{(i)}$ denote sorted magnitudes over all nonidentity Pauli strings. These sorted lists do not relabel the original coefficients $r_i$.

We use $\sigma_i$ for an indexed Pauli string and $\sigma_i^{\prime}$ when a distinct target string must be identified. Parentheses, as in $\mathbf Q=(\sigma'_1,\ldots,\sigma'_m)$, denote an ordered tuple: its $i$th entry is paired with $a_i$. Curly parantheses, as in $\mathcal Q=\{\sigma'_1,\ldots,\sigma'_m\}$, denote an unordered set.  

\section{One-Qubit Clifford Ergotropy}
\label{sec:one_qubit-ergo}

Analytically solving for the Clifford ergotropy requires tracking nontrivial Clifford conjugations across all parts of a Hamiltonian. For even modestly sized systems, this task becomes virtually infeasible; however, to demonstrate how one would track the conjugations, in principle, and turn them into an expression,  we first derive the general expression for the one qubit Clifford ergotropy. 

Any one-qubit Hamiltonian can be expanded in the basis of Pauli plus identity operators in the form $H=a_0\,I+\vec{a}\cdot\vec{\sigma}$, where $\vec a=(a_X,a_Y,a_Z)$, with indices $\alpha\in\{X,Y,Z\}$ and $\vec{\sigma}=(X,Y,Z)$.
Similarly, any one-qubit state can be represented as $\rho=\frac{1}{2}(I+\vec{r}\cdot\vec{\sigma})$. The Clifford minimum is therefore
\begin{equation}
E_{\mathrm{Cl}}^{\min}(\rho,H)
=
a_0
+
\min_{U\in\mathrm{Cl}_1}
 \sum_{\alpha\in\{X,Y,Z\}} a_\alpha\langle U^\dagger \sigma_\alpha U \rangle 
.
\end{equation}
The Clifford orbit of the one qubit Hamiltonian contains at most twenty-four distinct Hamiltonians, so in practice it is possible to enumerate every single permutation to calculate the minimum.

In any case, the average energy reads $\mtr{\rho H}=a_0+\vec r\cdot\vec a$. Since the identity contribution is invariant under conjugation, we can just optimize the term $\vec r\cdot\vec a$. A single-qubit Clifford operator acts on the Pauli matrices as a signed permutation,
\begin{equation}
    U\sigma_\alpha U^\dagger
    =
    s_\alpha\sigma_{\pi(\alpha)},
    \quad\text{with}\quad
    s_\alpha\in\{\pm1\}\,,
\end{equation}
where $ \pi\in S_3$. The corresponding signed-permutation matrix must preserve orientation and therefore satisfies
\begin{equation}
\label{eq:signs}
    \operatorname{sgn}(\pi)
    \prod_{\alpha\in\{X,Y,Z\}}s_\alpha
    =
    1.
\end{equation}
Here, $\operatorname{sgn}(\pi)=+1$ for an even permutation and $\operatorname{sgn}(\pi)=-1$ for an odd permutation. An even permutation necessitates an even number of sign flips, whereas an odd permutation allows an odd number of sign flips. For example, the odd permutation
$(a_X,a_Y,a_Z)\mapsto(a_X,a_Z,a_Y)$ has the four signed images
\begin{align}
    &(-a_X,a_Z,a_Y),\qquad
    (a_X,-a_Z,a_Y),\\ \nonumber
    &(a_X,a_Z,-a_Y),\qquad 
    (-a_X,-a_Z,-a_Y).
\end{align}
Each of the six unsigned permutations is associated with four allowed sign assignments, producing twenty-four signed transformations, which may yield coincident Hamiltonians. We  then formulate the Clifford minimum as an optimization over the six unsigned permutations. For a fixed permutation $\pi$, define $\lambda_\alpha^{(\pi)}=r_\alpha a_{\pi(\alpha)}$. If the signs could be selected independently, the minimum would be obtained by choosing
\begin{equation}
    s_\alpha^{(g)}
    =
    -\operatorname{sgn}
    \left(
        \lambda_\alpha^{(\pi)}
    \right),
\end{equation}
For $\lambda_\alpha^{(\pi)}=0$, choose either sign $s_\alpha^{(g)}\in\{\pm1\}$. Thus $s_\alpha^{(g)}\lambda_\alpha^{(\pi)}
=-|\lambda_\alpha^{(\pi)}|$. This gives the unconstrained, or greedy, minimum $
    -\sum_{\alpha\in\{X,Y,Z\}}
    \left|
        \lambda_\alpha^{(\pi)}
    \right|, $ which consequently is a lower bound on the Clifford minimum energy  as shown in Ref.~\cite{maity2026clifford}. 
    
If the greedy assignment satisfies Eq.~\eqref{eq:signs}, then it is in the Clifford orbit and is therefore the minimum associated with that permutation. Otherwise, one sign must be reversed. Reversing the $\alpha$th sign changes the corresponding contribution from
$-|\lambda_\alpha^{(\pi)}|$ to
$+|\lambda_\alpha^{(\pi)}|$, increasing the energy by
$2|\lambda_\alpha^{(\pi)}|$. The optimal correction is therefore to reverse the contribution with the smallest magnitude. Defining
\begin{equation}
    \chi_\pi
    =
    \frac{1}{2}
    \left[
        1-
        \operatorname{sgn}(\pi)
        \prod_{\alpha\in\{X,Y,Z\}}
        s_\alpha^{(g)}
    \right],
\end{equation}
the minimum associated with the unsigned permutation $\pi$ is
\begin{equation}
    E_{\mathrm{Cl},\pi}^{\min}
    =
    -\sum_{\alpha\in\{X,Y,Z\}}
    \left|
        r_\alpha a_{\pi(\alpha)}
    \right|
    +
    2\chi_\pi
    \min_{\alpha}
    \left|
        r_\alpha a_{\pi(\alpha)}
    \right|.
\end{equation}

Since only six unsigned permutations exist, the optimization reduces to six candidate minima. To write them explicitly, we define
\begin{equation}
    \eta
    =
    \operatorname{sgn}
    \left(
        r_Xr_Yr_Za_Xa_Ya_Z
    \right),
\end{equation}
together with
\begin{equation}
    \chi_{\mathrm e}
    =
    \frac{1+\eta}{2},
\quad\text{and}\quad
    \chi_{\mathrm o}
    =
    \frac{1-\eta}{2}\,.
\end{equation}
If any of the product vanishes, $\eta=0$ and both equal $\chi_{\mathrm e}=\chi_{\mathrm o}=1/2$. The three even-permutation candidates are
\begin{equation}
\begin{aligned}
E_{\mathrm{Cl},XYZ}^{\min}
&=
-|r_Xa_X|-|r_Ya_Y|-|r_Za_Z| \\
&\quad +2\chi_{\mathrm e}
\min\!\left\{
|r_Xa_X|,|r_Ya_Y|,|r_Za_Z|
\right\},
\\[1em]
E_{\mathrm{Cl},YZX}^{\min}
&=
-|r_Xa_Y|-|r_Ya_Z|-|r_Za_X| \\
&\quad +2\chi_{\mathrm e}
\min\!\left\{
|r_Xa_Y|,|r_Ya_Z|,|r_Za_X|
\right\},
\\[1em]
E_{\mathrm{Cl},ZXY}^{\min}
&=
-|r_Xa_Z|-|r_Ya_X|-|r_Za_Y| \\
&\quad +2\chi_{\mathrm e}
\min\!\left\{
|r_Xa_Z|,|r_Ya_X|,|r_Za_Y|
\right\}.
\end{aligned}
\end{equation}
The three odd-permutation candidates are
\begin{equation}
\begin{aligned}
E_{\mathrm{Cl},XZY}^{\min}
&=
-|r_Xa_X|-|r_Ya_Z|-|r_Za_Y| \\
&+2\chi_{\mathrm o}
\min\!\left\{
|r_Xa_X|,|r_Ya_Z|,|r_Za_Y|
\right\},
\\
E_{\mathrm{Cl},YXZ}^{\min}
&=
-|r_Xa_Y|-|r_Ya_X|-|r_Za_Z|\\
&+2\chi_{\mathrm o}
\min\!\left\{
|r_Xa_Y|,|r_Ya_X|,|r_Za_Z|
\right\},
\\
E_{\mathrm{Cl},ZYX}^{\min}
&=
-|r_Xa_Z|-|r_Ya_Y|-|r_Za_X|\\
&+2\chi_{\mathrm o}
\min\!\left\{
|r_Xa_Z|,|r_Ya_Y|,|r_Za_X|
\right\}.
\end{aligned}
\end{equation}
The single-qubit Clifford ergotropy is then
\begin{equation}
    \mathcal E_{\mathrm{Cl}}(\rho,H)
    =
    \vec r\cdot\vec a
    -
    \min_{\pi\in S_3}E_{\mathrm{Cl},\pi}^{\min}.
    \label{eq:one-qubit_ergo}
\end{equation}
To recap, solving the one-qubit Clifford ergotropy requires calculating the minimum energy for every unsigned permutation of a possible three entry Hamiltonian coefficient vector. Due to sign constraints, finding the permutation minimum requires sign flipping the smallest state-vector weighted term. In the limiting case where at least one Hamiltonian coefficient $a_\alpha$ or one Bloch-vector component $r_\beta$ vanishes, the Clifford optimization simplifies considerably. For every unsigned permutation, one of the matched products $r_\alpha a_{\pi(\alpha)}$ vanishes, meaning the parity-conserving sign flip occurs at no energy cost, and hence
\begin{equation}
 E_{\mathrm{Cl},\pi}^{\min}=-\sum_{\alpha\in\{X,Y,Z\}}\left|r_\alpha a_{\pi(\alpha)}\right|\,.   
\end{equation}
    
To simplify this further, let $A_{(1)}\ge A_{(2)}\ge A_{(3)}$ and $R_{(1)}\ge R_{(2)}\ge R_{(3)}$ be the independently sorted magnitudes of $\vec{a}$ and $\vec{r}$, respectively. By the rearrangement inequality, the largest value of $\sum_\alpha|r_\alpha a_{\pi(\alpha)}|$ pairs these magnitudes in descending order. The Clifford ergotropy in this limiting case is therefore
\begin{equation}
\mathcal E_{\mathrm{Cl}}(\rho,H)=\vec r\cdot\vec a+\sum_{i=1}^{3}A_{(i)}R_{(i)}.
\end{equation}
Here the third product vanishes under the stated assumption; the original vector components retain their Pauli-axis labels.
From a computational perspective, Eq.~\eqref{eq:one-qubit_ergo} offers little advantage over direct enumeration: evaluating the six permutation minima requires effort comparable to enumerating the full single-qubit Clifford orbit. Nevertheless, the derivation exposes the fundamental difficulty in obtaining analytic expressions for Clifford ergotropy. The allowed permutations and sign choices are constrained by the structure of Clifford conjugation, producing a discrete combinatorial optimization that becomes increasingly complicated as the system size grows. Simple closed-form expressions are generally available only when additional structure is imposed on the state or Hamiltonian. In the following sections, we identify several limiting classes of larger Hamiltonians which possess additional structure that substantially simplifies the optimization.

\section{Hamiltonians respecting one commutation relation}
\label{sec:one_commutation_relation}

In the preceding section, we have seen that evaluating the single qubit Clifford ergotropy required extensive bookkeeping of conjugation relations. For multi-qubit systems, these become effectively impossible to track.  Now, we continue by considering the class of Hamiltonians generated by a stabilizer set, and demonstrate how a relatively clean solution for the Clifford ergotropy can be found. We then generalize the discussion to Hamiltonians composed of anti-commuting generators and show that this class admits an equivalent form of the Clifford ergotropy.

\subsection{Stabilizer Hamiltonians}

The stabilizer formalism is used extensively throughout quantum information theory, particularly in the construction and analysis of quantum error-correcting codes
\cite{Gottesman1997,NielsenChuang2010}. An $n$-qubit stabilizer group is an Abelian subgroup of the $n$-qubit Pauli group
\begin{equation}
    \mathcal{S}
    =
    \left\langle
        S_1,\ldots,S_m
    \right\rangle < \mathcal{P}_n,
\end{equation}
where $m\le n$ and the generators satisfy
\begin{equation}
    [S_i,S_j]=0,
    \qquad
    S_i^2=I,
    \qquad
    -I\notin\mathcal{S}.
\end{equation}

We define a stabilizer Hamiltonian associated with an ordered tuple of independent stabilizer generators \(\mathbf{S}=(S_1,\ldots,S_m)\)
as
\begin{equation}
    H_{\mathbf{S}}
    =
    a_0I
    +
    \sum_{i=1}^{m} a_i S_i,
    \qquad
    a_i\in\mathbb{R}.
    \label{eq:stabilizer_hamiltonian}
\end{equation}
Stabilizer Hamiltonians appear naturally in quantum error correction, where commuting Pauli operators define the stabilizer checks of a quantum code, and in exactly solvable models of many-body physics, including the toric code and cluster-state Hamiltonians. More generally, they serve as commuting-projector parent Hamiltonians whose ground spaces are stabilizer states or stabilizer-code subspaces \cite{Gottesman1997,kitaev2003fault,raussendorf2001oneway}. Their physical relevance and highly constrained mathematical structure make them a natural starting class of Hamiltonians for which to study the Clifford ergotropy minimum. 

Let $\operatorname{Stab}_m(\mathcal{P}_n)$ denote the set of ordered $m$-tuples of independent, mutually commuting signed Hermitian Pauli operators. Let $\operatorname{Stab}_m^+(\mathcal{P}_n)$ denote its unsigned counterpart, with every entry in $\mathcal P_n^*$,
\begin{equation}
    \operatorname{Stab}_m(\mathcal{P}_n)
    \equiv
    \left\{(\sigma'_1,\ldots,\sigma'_m) : \sigma'_i\in\pm\mathcal P_n^*,\ [\sigma'_i,\sigma'_j]=0\right\}.
\end{equation}
where we need the $(\sigma'_1,\ldots,\sigma'_m)$ to be independent from each other. Since Clifford conjugation preserves the Pauli group, commutation
relations, and independence, the Clifford orbit of
$H_{\mathbf{S}}$ is
\begin{equation}
    \mathcal{O}_{\mathrm{Cl}}(H_{\mathbf{S}})
    =
    \left\{
        a_0I
        +
        \sum_{i=1}^{m}a_i \sigma'_i : 
        (\sigma'_1,\ldots,\sigma'_m)
        \in\operatorname{Stab}_m(\mathcal{P}_n)
    \right\}.
    \label{eq:stabilizer_clifford_orbit}
\end{equation}
Therefore, the Clifford
minimum energy reduces to an optimization over all ordered stabilizer-generator tuples,
\begin{equation}
    E_{\mathrm{Cl}}^{\min}(\rho,H_{\mathbf{S}})
    =
    a_0+
    \min_{\mathbf{Q}\in\operatorname{Stab}_m(\mathcal{P}_n)}
    \sum_{i=1}^{m}a_i\mtr{\rho \sigma'_i}.
    \label{eq:stabilizer_clifford_minimum}
\end{equation}
At first glance, Eq.~\eqref{eq:stabilizer_clifford_minimum} gives only a modest advantage when compared to enumerating over every Clifford operation. The number of ordered independent commuting $m$-tuples of unsigned Pauli operators (see appendix \ref{app:stabilizer_count}) is
\begin{equation}
    \left|
\operatorname{Stab}_m^+(\mathcal{P}_n)
    \right|
    =
    \prod_{k=0}^{m-1}
    \left(2^{2n-k}-2^k\right),
    \label{eq:number_ordered_stabilizer_sets}
\end{equation}
which is exponential in the system size. However, the
structure of a stabilizer set allows for further simplification. We will see shortly in Theorem~\ref{thm:independent_sign_freedom} that the signs of an independent set of stabilizer
generators may be chosen independently under Clifford conjugation. That is, for every choice $\boldsymbol{\epsilon}=(\epsilon_1,\ldots,\epsilon_m)$, with $\epsilon_i\in\{\pm1\}$ chosen independently of the Hamiltonian coefficient $a_i$, there exists a Clifford unitary $U$ such that $ U S_i U^\dagger=\epsilon_i \sigma'_i$, for any ordered independent commuting Pauli tuple $(\sigma'_1,\ldots,\sigma'_m)$. Using this, let $\mathbf{a}=(a_1,\ldots,a_m)$ denote the ordered Hamiltonian coefficients, where $a_i$ is associated with the stabilizer generator $S_i$. We expand the input state in the Pauli basis according to Eq.~\eqref{eq:pauli_expansions}. Then, without loss of generality, reorder the Hamiltonian generators and coefficients such that
\begin{equation}
    |a_1|\geq |a_2|\geq\cdots\geq |a_m|.
\label{eq:ordered_hamiltonian_coefficients}
\end{equation}
Let $\mathfrak C_m$ be the family of unordered $m$-element subsets of $\mathcal P_n^*$ that are independent and mutually commuting. For each $\mathcal Q=\{\sigma'_1,\ldots,\sigma'_m\}\in\mathfrak C_m$, let
\begin{equation}
    r_{(1)}(\mathcal{Q})
    \geq
    r_{(2)}(\mathcal{Q})
    \geq\cdots\geq
    r_{(m)}(\mathcal{Q})
\end{equation}
denote the decreasing rearrangement of the expectation-value magnitudes $\left\{|\mtr{\rho \sigma'_i}|:\sigma'_i\in\mathcal{Q}\right\}. $ Since every permutation and every independent choice of generator
signs is accessible under Clifford conjugation, the rearrangement
inequality implies that the largest Hamiltonian coefficient should be
paired with the largest allowed Pauli expectation value. Therefore,
\begin{equation}
    E_{\mathrm{Cl}}^{\min}(\rho,H_{\mathbf S})
    =
    a_0-
    \max_{\substack{
        \mathcal{Q}\in\mathfrak C_m
    }}
    \sum_{i=1}^{m}
    |a_i|\,r_{(i)}(\mathcal{Q}).
    \label{eq:stabilizer_minimum_sorted}
\end{equation}
Computationally, evaluating Eq.~\eqref{eq:stabilizer_minimum_sorted} requires searching over every valid $m$-element stabilizer set, or equally every independent mutually commuting subset of $\mathcal{P}_n^H$. Since the ordering within each set is removed by sorting its Pauli expectation values, the number of sets that must be considered are
\begin{equation}
    N_{n,m}
    =
    \frac{1}{m!}
    \prod_{k=0}^{m-1}
    \left(2^{2n-k}-2^k\right).
    \label{eq:number_valid_stabilizer_sets}
\end{equation}
Sorting removes the need to test the $m!$ orderings of each unordered set, while independent sign freedom removes explicit sign enumeration. Equation~\eqref{eq:number_ordered_stabilizer_sets} counts unsigned ordered tuples, not distinct signed Hamiltonians. Coefficient symmetries can make different tuples represent the same Hamiltonian. For fixed $m$, the number of candidate sets still grows exponentially with $n$.

For certain input states, the optimization simplifies further. Let $\rho=\ket{\psi}\!\bra{\psi}$ be a pure stabilizer state with stabilizer group $\mathcal{S}_{\psi}$. Its Pauli expectation values
satisfy
\begin{equation}
    \left|\mtr{\rho \sigma_i}\right|
    =
    \begin{cases}
        1, & \sigma_i\in\pm\mathcal{S}_{\psi},\\
        0, & \sigma_i\notin\pm\mathcal{S}_{\psi}.
    \end{cases}
\end{equation}
Since $\mathcal{S}_{\psi}$ is composed of or generates a full stabilizer set, there must exist some Clifford transformation which maps the Pauli elements in the system Hamiltonian to elements in $\mathcal{S}_{\psi}$, meaning that there exists a stabilizer set in the Clifford
orbit of the original Hamiltonian whose generators stabilize $\ket{\psi}$. Independent sign freedom then allows every contribution
to be made negative, giving
\begin{equation}
    E_{\mathrm{Cl}}^{\min}(\rho,H_{\mathbf S})
    =
    a_0-\sum_{i=1}^{m}|a_i|.
    \label{eq:stab_state_minimum}
\end{equation}
For $m=n$, the selected generators may be chosen to generate the full stabilizer group of the input state.

Stabilizer states are a special case of a larger simplification. Suppose the input state expanded in the Pauli basis has its $m$ largest nonidentity coefficient magnitudes on an independent, mutually commuting unordered Pauli set. Using the logic above, those $m$ Pauli strings are part of a stabilizer set. Through sign freedom, any signed permutation exists in the Clifford orbit. The globally sorted magnitudes defined in Sec.~\ref{sec:definitions} then satisfy
\begin{equation}
    R_{(1)}
    \geq
    R_{(2)}
    \geq\cdots\geq
    R_{(m)},
\end{equation}
and pair them with the decreasingly ordered Hamiltonian coefficients
\begin{equation}
    |a_1|\geq |a_2|\geq\cdots\geq |a_m|.
\end{equation}
Since these are the globally largest Pauli coefficients of the input
state, the assignment gives the upper bound on the Clifford ergotropy, so no searching is required, and
\begin{equation}
    E_{\mathrm{Cl}}^{\min}(\rho,H_{\mathbf S})
    =
    a_0-
    \sum_{i=1}^{m}
    |a_i|\,R_{(i)}.
    \label{eq:commuting_largest_weights_minimum}
\end{equation}

\subsection{Pairwise Anticommuting Hamiltonians}
 
In the preceding subsection, we showed that restricting a Hamiltonian to an independent mutually commuting set of Pauli operators substantially simplifies its Clifford orbit and, consequently, the calculation of its Clifford ergotropy. The derivation relied on two properties. First, independent sign freedom removes the need to search over the $2^m$ possible sign assignments. Second, because every permutation of a mutually commuting set preserves its commutation relations, permutation freedom makes the $m!$ ordered realizations of each set redundant.

It is then natural to consider the analogous class of Hamiltonians whose Pauli terms are pairwise anticommuting. Such Hamiltonians arise naturally in Dirac-type models, whose gamma matrices generate a Clifford algebra \cite{dirac1928quantum}, and in fermionic systems through Majorana operators satisfying canonical anticommutation relations \cite{elliott2015majorana}. Anticommuting Pauli families also appear in quantum-simulation algorithms, where they are used to reduce the measurement cost of variational energy estimation \cite{izmaylov2020unitary}.

We define an independent pairwise anticommuting Hamiltonian by
\begin{equation}
    H
    =
    a_0I
    +
    \sum_{i=1}^{m}a_i\sigma_i,
    \qquad
    \sigma_i\in\mathcal{P}_n^{H},
\label{eq:anticommuting_hamiltonian}
\end{equation}
where $\{\sigma_i,\sigma_j\}=2\delta_{ij}I$, and the ordered tuple $(\sigma_1,\ldots,\sigma_m)$ is additionally assumed independent modulo phase. The identity $\sigma_i^2=I$ alone does not imply independence. By Theorems~\ref{thm:pauli_set_orbit} and~\ref{thm:independent_sign_freedom}, any two such ordered independent tuples are connected by Clifford conjugation, with arbitrary signs. The signed Hamiltonian orbit is therefore
\begin{equation}
    \mathcal{O}_{\mathrm{Cl}}(H)
    =
    \left\{
        a_0I
        +
        \sum_{i=1}^{m}s_i a_i\sigma'_i : \sigma'_i\in\mathcal P_n^*,\ s_i\in\{\pm1\}\right\}\,,\label{eq:anticommuting_clifford_orbit}
\end{equation}
where now $\{\sigma'_i,\sigma'_j\}=2\delta_{ij}I$ and again the $(\sigma'_1,\ldots,\sigma'_m)$ are independent.
Then, by exactly the same logic as in the commuting case, the Clifford minimum is found by searching over every unsigned independent pairwise anticommuting Pauli set. Let
\begin{equation}
    |a_1|\geq |a_2|\geq\cdots\geq |a_m|,
\end{equation}
and let $\mathfrak A_m$ be the family of unordered, independent, pairwise anticommuting $m$-element subsets of $\mathcal P_n^*$. For each $\mathcal Q=\{\sigma'_1,\ldots,\sigma'_m\}\in\mathfrak A_m$, let
\begin{equation}
    r_{(1)}(\mathcal{Q})
    \geq
    r_{(2)}(\mathcal{Q})
    \geq\cdots\geq
    r_{(m)}(\mathcal{Q})
\end{equation}
denote the decreasingly ordered expectation-value magnitudes of the set $\left\{|\mtr{\rho \sigma'_i}|:\sigma'_i\in\mathcal{Q} \right\}$.
Using the same sign freedom argument as the commuting case, the rearrangement inequality gives
\begin{equation}
    E_{\mathrm{Cl}}^{\min}(\rho,H)
    =
    a_0-
    \max_{\mathcal{Q}\in\mathfrak A_m}
    \sum_{i=1}^{m}
    |a_i|\,r_{(i)}(\mathcal{Q}).
    \label{eq:anticommuting_clifford_minimum}
\end{equation}
As in the commuting case, the explicit search over signs and ordered permutations is replaced by a weighted search over unordered Pauli sets with the required commutation structure. In a similar vein, consider an input state whose $m$-largest Pauli-basis expanded weights lie on an independent, pairwise anticommuting unordered set. Then, by the same logic as in the commuting case, the Clifford minimum is identical to Eq.~\eqref{eq:commuting_largest_weights_minimum}.

To recap, imposing strict commutation structure -- mutually commuting or anti-commuting -- greatly simplifies the computational cost in evaluating the Clifford ergotropy. The gain stems from the sign and permutation freedom granted by Clifford conjugation preserving commutation relations, the Pauli group, and, most importantly, independence, which will be discussed in the next section. Furthermore, for input states whose largest Pauli-basis weights matched the commutation structure, the Clifford minimum saturates its upper bound. 

It is therefore natural to ask how these results extend to
Hamiltonians with mixed commutation structure. In this setting, we cannot generally say that every permutation or sign choice exists in the Clifford orbit. The question then is what structure must we impose such that some simplification emerges. The next section will discuss a class of Hamiltonians allowing for a mixed commutation structure while retaining a simplified Clifford orbit.

\section{Clifford Orbits of Linearly Independent Pauli Hamiltonians}
\label{sec:binary_symplectic}

In the prior sections, we discussed how sign freedom allows for measurable simplification in computing the Clifford ergotropy, but gave no reasoning as to why sign freedom was guaranteed. Furthermore, in discussing the commuting and anti-commuting cases, we restricted the number of allowed Pauli-elements such that the list remained independent. It turns out, to motivate these decisions, it is necessary to introduce the symplectic group underlying the Clifford group. In this section, we first introduce the symplectic binary group, then  relay the fundamental theorems underpinning this entire work, and follow with a broad class of Hamiltonians for which calculating the Clifford ergotropy is reasonable: binary symplectic group independent Hamiltonians. 

\subsection{Binary Symplectic Representation}
The binary symplectic representation establishes an isomorphism between the $n$-qubit Pauli group modulo its phase factors and the vector space $\mathbb{F}_2^{2n}$ \cite{DehaeneDeMoor2003,Gottesman1997}. Under the convention
\begin{equation}
    \sigma(v)
    =
    i^{\kappa(v)}
    \bigotimes_{j=1}^{n}
    X_j^{x_j}Z_j^{z_j},
    \qquad
    v
    =
    (x\mid z)
    \in\mathbb{F}_2^{2n},
    \label{eq:binary_pauli_map}
\end{equation}
the single-qubit Pauli operators are represented by
\begin{align}
    &I\longleftrightarrow (0\mid0),
    &
    X\longleftrightarrow (1\mid0),\\
    & Z\longleftrightarrow (0\mid1),
    &Y\longleftrightarrow (1\mid1).
    \label{eq:single_qubit_binary_paulis}
\end{align}
We choose $\kappa(v)=\sum_j x_jz_j$ (an integer sum), so $\sigma(v)$ is the unsigned Hermitian Pauli string. When a Pauli string is specified by its binary vector $v$, we write $\sigma(v)$ and $r_v:=\mtr{\rho\sigma(v)}$. In particular, $\sigma_i=\sigma(v_i)$ and $r_i=r_{v_i}$ for an indexed Pauli tuple. All binary vectors are column vectors; the notation $v=(x\mid z)$ lists their two blocks. Multiplication
of Pauli operators modulo phase corresponds to binary vector addition,
\begin{equation}
    \sigma(v)\sigma(w)
    \sim
    \sigma(v+w),
\end{equation}
up to phase factors in $\{\pm1,\pm i\}$. A set of vectors $\{v_1,v_2,\ldots,v_m\} \subset\mathbb{F}_2^{2n}$ is linearly independent over $\mathbb{F}_2$,
\begin{equation}
    \sum_{i=1}^{m}c_iv_i=\mathbf{0},
    \qquad
    c_i\in\mathbb{F}_2,
\end{equation}
only if $c_1=c_2=\cdots=c_m=0. $ Note that independence of Pauli strings means independence modulo phase: no nonempty product is proportional to $I$, equivalently their binary vectors are linearly independent over $\mathbb F_2$.

Although symplectic geometry constitutes a broad field of study, for the purposes of this work we require only its basic definitions. A symplectic group is a group of invertible linear transformations that preserve a nondegenerate alternating bilinear form. In the binary Pauli representation, the relevant form encodes the commutation relations between Pauli operators. For $v=(x\mid z) $ and $w=(x'\mid z')$
we define the binary symplectic form by
\begin{equation}
    \omega(v,w)
    =
    x\cdot z'
    +
    z\cdot x'
    \pmod 2,
    \label{eq:binary_symplectic_form}
\end{equation}
which encodes Pauli commutation according to
\begin{align}
    \omega(v,w)=0
    &\iff
    [\sigma(v),\sigma(w)]=0,
    \\
    \omega(v,w)=1
    &\iff
    \{\sigma(v),\sigma(w)\}=0.
\end{align}
Upon fixing the standard binary Pauli basis, the symplectic form may
be written as
\begin{equation}
    \omega(v,w)
    =
    v^{T}Jw,
    \qquad
    J=
    \begin{pmatrix}
        0&I_n\\
        I_n&0
    \end{pmatrix},
    \label{eq:standard_symplectic_matrix}
\end{equation}
where $I_n$ is the $n\times n$ identity matrix. A symplectic matrix satisfies $S^TJS=J$, or equivalently $\omega(Sv,Sw)=\omega(v,w)$. Clifford conjugation induces such a binary transformation, and every $S\in\operatorname{Sp}(2n,2)$ has a Clifford representative; the binary action omits Pauli signs.

For an ordered tuple $\mathbf v=(v_1,\ldots,v_m)$ of binary vectors, define its commutation matrix by
\begin{equation}
\Omega(\mathbf v)_{ij}:=\omega(v_i,v_j).
\label{eq:commutation_matrix}
\end{equation}
Thus $v_i$ is a single column vector, while $\mathbf v$ is an ordered tuple of such vectors. Reordering a tuple permutes the corresponding rows and columns of $\Omega$ and must carry the Hamiltonian coefficients along with the Pauli terms.

\subsection{Linearly Independent Mixed Hamiltonians}

Consider Hamiltonians of the form
\begin{equation}
    H=a_0I+\sum_{i=1}^{m} a_i \sigma(v_i),
\end{equation}
where  $\sigma(v_i)\in \mathcal P_n^*$, $v_i\in \mathbb F_2^{2n}$, and the binary vectors \(\{v_i\}_{i=1}^{m}\) are linearly independent over
\(\mathbb F_2\). Since there are at maximum $2n$ independent binary vectors for a $n$-qubit system $m\le 2n$. Independence does not imply commutativity. For example $IZ$ and $IX$ are independent and do not commute, while $IZ$ and $XI$ are independent and do commute. Therefore the Hamiltonians considered here can have a mixed commutation structure. For this class of Hamiltonians, we prove that the Clifford orbit is 
\begin{equation}
\mathcal{O}_{\mathrm{Cl}}(H)
=
\left\{
a_0I
+
\sum_{i=1}^{m}s_i a_i \sigma(w_i)
\;:\;
\begin{array}{l}
\vec{w}=(w_1,\ldots,w_m)
\,\\
\Omega(\vec{w})=\Omega(\vec{v}),\\
s_i\in\{\pm1\}
\end{array}
\right\},
\label{eq:independent_hamiltonian_clifford_orbit}
\end{equation}
where every $\vec{w}$ is independent over $\mathbb F_2$. This characterization follows from the two theorems below and Corollary~\ref{cor:independent_hamiltonian_orbit}.

\begin{theorem}
\label{thm:pauli_set_orbit}
Let $\vec{v}=(v_1,\ldots,v_m)$ and $\vec{w}=(w_1,\ldots,w_m)$ be two ordered, $\mathbb{F}_2$-linearly independent tuples in $\mathbb{F}_2^{2n}$. Then there exists a symplectic transformation $S \in\operatorname{Sp}(2n,2)$ such that $Sv_i=w_i$ for all $i=1,\ldots,m$ if and only if $\omega(v_i,v_j)=\omega(w_i,w_j)$ for all $i,j=1,\ldots,m$.
\end{theorem}

The proof is given in Appendix~\ref{app:tuple_orbit_proof}.
The action of every $U\in\mathrm{Cl}_n$ on Pauli vectors is represented by some $S\in\operatorname{Sp}(2n,2)$, and every $S\in\operatorname{Sp}(2n,2)$ is induced by a Clifford operator up to Pauli phases. It follows from Theorem~\ref{thm:pauli_set_orbit} that the unsigned
Clifford orbit of an independent ordered tuple is
\begin{equation}
\mathcal{O}_{\mathrm{Cl}}^{+}(\vec{v})
=
\left\{
\vec{w}=(w_1,\ldots,w_m)
\in(\mathbb{F}_2^{2n})^m:\Omega(\vec{w})=\Omega(\vec{v})
\right\},
\label{eq:unsigned_independent_tuple_orbit}
\end{equation}
and $\vec{w}$ is independent over $\mathbb{F}_2$. The unsigned orbit does not retain the signs of the conjugated Pauli operators. The following result supplies the missing sign
information.

\begin{theorem}[Independent sign freedom] 
\label{thm:independent_sign_freedom}
Let $ \vec{w}=(w_1,\ldots,w_m)$ be an $\mathbb{F}_2$-linearly independent tuple in $\mathbb{F}_2^{2n}$. Then, for every sign assignment
\begin{equation}
    (s_1,\ldots,s_m)\in\{\pm1\}^m,
\end{equation}
there exists a Pauli operator $\sigma(u)$ for some $u\in \mathbb{F}_2^{2n}$ such that
\begin{equation}
    \sigma(u)\sigma(w_i)\sigma(u)^\dagger
    =
    s_i\sigma(w_i)
\end{equation}
for every $i=1,\ldots,m$.
\end{theorem}

The proof is given in Appendix~\ref{app:sign_freedom_proof}.
\begin{corollary}[Clifford orbit of an independent Pauli Hamiltonian]
\label{cor:independent_hamiltonian_orbit}
Let
\begin{equation}
    H
    =
    a_0I
    +
    \sum_{i=1}^{m}a_i\sigma(v_i),
\end{equation}
where \(\vec{v}=(v_1,\ldots,v_m) \) is $\mathbb{F}_2$-linearly independent. Then its Clifford orbit is given by Eq.~\eqref{eq:independent_hamiltonian_clifford_orbit}.
\end{corollary}

The proof is given in Appendix~\ref{app:hamiltonian_orbit_proof}.
It is of note that the commuting and anti-commuting Hamiltonians considered in section~\ref{sec:one_commutation_relation} are limiting cases of Eq.~\eqref{eq:independent_hamiltonian_clifford_orbit}, making independent Hamiltonians the most general case we found where some simplification of the orbit is obtained. 

\section{Evaluating the Ergotropy}
\label{sec:evaluating_ergo}
Equation~\eqref{eq:independent_hamiltonian_clifford_orbit} reduces the calculation to a search over ordered, independent unsigned Pauli tuples with the prescribed commutation matrix. Each position remains paired with its Hamiltonian coefficient, and sign optimization is performed analytically. We first count these unsigned tuples, then discuss exact optimization and an approximation scheme.

\subsection{Asymptotic Growth of the Linearly Independent Hamiltonian Clifford Orbit}
\label{subsec:growth_of_mixed}
We count the unsigned ordered-tuple orbit $\mathcal O_{\mathrm{Cl}}^+(\mathbf v)$ defined in Eq.~\eqref{eq:unsigned_independent_tuple_orbit}. Independent sign freedom supplies $2^m$ signed tuples for each unsigned tuple. Their images cover the Hamiltonian orbit, but coefficient symmetries can make distinct signed tuples represent the same Hamiltonian. Consequently, the signed-tuple count is an upper bound on the number of distinct Hamiltonians, with equality when the parametrization is injective.

Within this context, let $\vec{v}=(v_1,\ldots,v_m)$ be the $\mathbb{F}_2^{2n}$ representation of the Hamiltonian. Given the ordered prefix $V_k=(v_1,\ldots,v_k)$, assume the ordered prefix $W_k=(w_1,\ldots,w_k)$ was chosen such that all entries are linearly independent and satisfy the same commutation relations as $V_k$. Suppose we append the vector $v_{k+1}$. We ask: how many distinct choices of $w_{k+1}$ can be made such that the appended tuple $W_{k+1}=(w_1,\ldots,w_k,w_{k+1})$ remains linearly independent and satisfies the commutation relations $\omega(v_{k+1},v_i)=\omega(w_{k+1},w_i)$ for $i=1,\ldots,k$, as required by the Eq.~\eqref{eq:unsigned_independent_tuple_orbit}? To answer this question, we define the linear system
\begin{equation}
    L_k x=b_k,
\end{equation}
where
\begin{equation}
L_k=
\begin{pmatrix}
w_1^{T}J\\
w_2^{T}J\\
\vdots\\
w_k^{T}J
\end{pmatrix},
\qquad
b_k=
\begin{pmatrix}
\omega(v_1,v_{k+1})\\
\omega(v_2,v_{k+1})\\
\vdots\\
\omega(v_k,v_{k+1})
\end{pmatrix},
\end{equation}
and $J$ is the standard symplectic form. Here, $L_k\in\mathbb{F}_2^{k\times 2n}$ is constructed from the previously chosen vectors $w_1,\ldots,w_k$, while $x\in\mathbb{F}_2^{2n}$ is the binary symplectic vector corresponding to a candidate Pauli operator $\sigma(x)$. The equation $L_kx=b_k$ imposes $ \omega(w_i,x)=\omega(v_i,v_{k+1}),$ for $i=1,\ldots,k$, ensuring that $\sigma(x)$ has the required commutation relations with each of the previously chosen Pauli operators $\sigma(w_1),\ldots,\sigma(w_k)$. Since the rows of $L_k$ are linearly independent, $\operatorname{rank}(L_k)=k$. Then, by the rank--nullity theorem, $\dim\ker(L_k)=2n-k$. Since $L_k: \mathbb F_{2}^{2n}\mapsto \mathbb F_{2}^{k}$ and is rank $k$, the map is surjective, there must exist some solution to the linear equation. If one particular solution is $x_0$, the solution set is given by the affine space $\mathcal{A}=\{x_0+a:a \in \ker(L_k), b_k=L_kx_0 \}$. Since $\ker(L_k)$ is a $(2n-k)$-dimensional vector space over $\mathbb{F}_2$, it contains $ \left|\ker(L_k)\right|=2^{2n-k}$ elements. Translation by $x_0$ is a bijection from $\ker(L_k)$ onto $x_0+\ker(L_k)$, so $|\mathcal{A}|=2^{2n-k}$. There are therefore $2^{2n-k}$ Pauli vectors satisfying the inhomogeneous system before the linear-independence condition is imposed.

We then recursively enumerate $|\mathcal{O}_{\mathrm{Cl}}^{+}(\vec{v})|$. Let

\begin{equation}
    N_k
    =
    2^{2n-k}
    -
    \left|
    \left\{
    x\in\operatorname{span}(W_k)
    :
    L_kx=b_k
    \right\}
    \right|,
\end{equation}
where $N_k$ denotes the number of unique choices of $w_{k+1}$ after fixing $w_1,\ldots,w_k$. The first term counts all solutions of $L_kx=b_k$, while the second removes those already contained in $\operatorname{span}(W_k)$, which would violate linear independence. Then
\begin{equation}
    |\mathcal{O}_{\mathrm{Cl}}^{+}(\vec{v})|
    =
    (2^{2n}-1)
    \prod_{k=1}^{m-1}N_k.
\end{equation}

The prefactor counts the possible nonzero choices of $w_1$. Two limiting cases exist for $N_k$--mutually commuting and mutually anti-commuting Hamiltonians. First, if the appended vector $v_{k+1}$ commutes with every element of $V_k=(v_1,\ldots,v_k)$, then $b_k=0$. If, in addition, $W_k$ is mutually commuting, every vector in $\operatorname{span}(W_k)$ is a solution of the $k$th linear system. Therefore, $N_k=2^{2n-k}-2^k$. Second, if no vector in $\operatorname{span}(W_k)$ satisfies the $k$th linear system, then no vectors are removed from the solution set. For $1\leq k\leq m-1$, the preceding cases give
\begin{equation}
    2^{2n-k}-2^k
    \leq N_k
    \leq 2^{2n-k}.
\end{equation}
The lower bound is strictly positive when $k<n$, yielding a clean lower bound when $m\leq n$. The restriction stems from there existing only $n$ independent commuting Pauli tuples in $\mathbb F_{2}^{2n}$, so when $k=n$ no additional vector can be appended such that the commutation and independence conditions are preserved. For $k>n$, the set
$W_k$ itself cannot be both mutually commuting and linearly
independent. The size of the orbit is then bounded by
\begin{equation}
2^{\,2nm-\frac{m(m-1)}{2}}
\prod_{j=n-m+1}^{n}
\left(1-4^{-j}\right)
\leq |\mathcal{O}_{\mathrm{Cl}}(\mathbf{v})|
\end{equation} 
and
\begin{equation}
|\mathcal{O}_{\mathrm{Cl}}^{+}(\vec{v})|\le
(2^{2n}-1)
2^{\,2n(m-1)-m(m-1)/2}\,.
\end{equation}
The asymptotic growth of the unsigned Clifford orbit is therefore
\begin{equation}
|\mathcal{O}_{\mathrm{Cl}}^{+}(\vec{v})|
=
\mathcal{O}\!\left(
2^{\,2nm-\frac{m(m-1)}{2}}
\right).
\end{equation}
Including the independent sign choices, the size of the Clifford orbit
of an independent Pauli Hamiltonian grows as
\begin{equation}
    \left|\mathcal{O}_{\mathrm{Cl}}(H)\right|
    =
    \mathcal{O}\!\left(
        2^{\,2nm-\frac{m(m-3)}{2}}
    \right).
\end{equation}
As remarked at the beginning of this subsection, calculating the Clifford minimum by exhaustive search requires minimizing over every element of the unsigned orbit. Assuming that the objective function can be evaluated efficiently for each orbit element, the computational cost is therefore proportional to the size of the searchable set. The value of restricting to independent Pauli Hamiltonians is now clear: when $m$ is fixed, the super-exponential growth with the number of qubits exhibited by the full Clifford group in Eq.~\eqref{eq:clifford_group_size} is reduced to exponential growth. However it is necessary to mention that when $m \rightarrow n$, the super exponential scaling returns as the orbit encompass the full action of the Clifford group. We also emphasize that $m\le 2n$ to ensure linear independence, 
Thus, linear independence of the Pauli terms produces a substantial
reduction in the complexity of the exhaustive Clifford-ergotropy
calculation.

\subsection{Calculating the mixed Hamiltonian Clifford Ergotropy}
\label{subsec:calc_ergo}

We conclude the analysis with a method of calculating the Clifford ergotropy by a non-exhaustive search, and an approximation scheme polynomial in the supplied Pauli-expansion size. 

\paragraph*{Exact calculation}
Starting from the Clifford minimum given in Eq.~\eqref{eq:hamiltonian_clifford_minimum}, we expand the Hamiltonian to obtain
\begin{equation}
E_{\mathrm{Cl}}^{\min}(\rho,H)
=a_0+
\min_{\substack{
\vec{w}\in\mathcal{O}_{\mathrm{Cl}}^{+}(\vec{v})\\
\mathbf{s}\in\{\pm1\}^{m}
}}
\sum_{i=1}^{m}
a_i s_i\mtr{\rho \sigma(w_i)}.
\end{equation}
Thus, every element of the Clifford orbit selects $m$ distinct Pauli expectation values from the input state and weights them by the Hamiltonian coefficients. Because the Pauli vectors defining the Hamiltonian are linearly independent, the signs may be chosen independently. Consequently,
\begin{equation}
E_{\mathrm{Cl}}^{\min}(\rho,H)
=
a_0-
\max_{\vec{w}\in
\mathcal{O}_{\mathrm{Cl}}^{+}(\vec{v})}
\sum_{i=1}^{m}
|a_i|
\left|
\mtr{\rho \sigma(w_i)}
\right|.
\end{equation}

To solve this optimization, we first compute and sort the absolute Pauli expectation values
\begin{equation}
    \left|\mtr{\rho \sigma(w)}\right|,
    \qquad
    w\in\mathbb{F}_2^{2n}\setminus\{0\},
\end{equation}
while retaining the Pauli vector associated with each value. If the Hamiltonian is evenly weighted, i.e., $|a_1|=|a_2|=\cdots=|a_m|, $ then the optimization reduces to finding the $m$ Pauli expectation values having the largest total magnitude whose corresponding vectors are linearly independent and reproduce the commutation matrix of the Hamiltonian.

For unequal coefficient magnitudes, this is a constrained weighted-assignment problem over ordered tuples. Reorder the Hamiltonian terms, carrying their binary vectors along, so that $|a_1|\ge\cdots\ge|a_m|$. Let $u_i$ label the Pauli string attaining the $i$th globally sorted magnitude $R_{(i)}=|r_{u_i}|$. If the ordered tuple $(u_1,\ldots,u_m)$ is independent and satisfies $\Omega(u_1,\ldots,u_m)=\Omega(\mathbf v)$, the rearrangement bound is attained,
\begin{equation}
E_{\mathrm{Cl}}^{\min}(\rho,H)=a_0-\sum_{i=1}^{m}|a_i|R_{(i)}.
\end{equation}
The ordering matters: the existence of some commutation-compatible ordering of the same unordered set does not by itself guarantee that the largest state and Hamiltonian magnitudes can be paired.

In the general case, the structure of the orbit lends itself naturally to a graph representation in which the commutation matrix defines a commutation graph. The optimization may then be viewed as a maximum-weight induced subgraph-matching problem with an additional linear-independence constraint. Strategies such as branch-and-bound, constraint propagation, and symmetry reduction may be used to search for the Clifford minimum \cite{ullmann1976algorithm,cordella2004subgraph,schrijver2003combinatorial}. Although these methods may substantially reduce the practical runtime, the worst-case search space remains exponential in the number of qubits. This motivates the use of approximate methods.

\paragraph*{Clifford Minimum Approximation}

Using the linear construction of Sec.~\ref{subsec:growth_of_mixed}, we may construct an approximation of the Clifford ergotropy. Let
\begin{equation}
H=a_0I+\sum_{i=1}^{m}a_i\sigma(v_i),
\qquad
r_w=\tr{\rho \sigma(w)},
\end{equation}
where the vectors $v_1,\ldots,v_m\in\mathbb F_2^{2n}$ are linearly independent. We first reorder the Hamiltonian terms, carrying the Pauli vectors along with their coefficients, such that
\begin{equation}
|a_1|\ge |a_2|\ge\cdots\ge |a_m|.
\end{equation}
The exact Clifford optimization may be written as
\begin{equation}
M_{\Omega}(\rho,\mathbf a)
=
\max_{\substack{
\Omega(\mathbf w)=\Omega(\mathbf v)\\
\operatorname{rank}(w_1,\ldots,w_m)=m
}}
\sum_{i=1}^{m}|a_i|\,|r_{w_i}|,
\end{equation}
where the absolute values arise from independent sign freedom guaranteed by theorem \ref{thm:independent_sign_freedom}. To make an approximation, instead of searching the entire Clifford orbit for the optimal Pauli set, we construct an admissible tuple one vector at a time. Suppose the procedure has already selected the independent Pauli vectors $w_1^{\mathrm g},\ldots,w_{k-1}^{\mathrm g}$. Define
\begin{equation}
L_{k-1}
=
\begin{pmatrix}
(w_1^{\mathrm g})^TJ\\
(w_2^{\mathrm g})^TJ\\
\vdots\\
(w_{k-1}^{\mathrm g})^TJ
\end{pmatrix},
\qquad
b_k
=
\begin{pmatrix}
\omega(v_1,v_k)\\
\omega(v_2,v_k)\\
\vdots\\
\omega(v_{k-1},v_k)
\end{pmatrix}.
\end{equation}
The commutation constraints on the next vector are precisely the affine linear system $L_{k-1}w=b_k$ over $\mathbb F_2$. The feasible set of allowed vectors at the $k$th step is therefore
\begin{equation}
\mathcal F_k
=
\left\{
w\in\mathbb F_2^{2n}:
L_{k-1}w=b_k,\,
w\notin\operatorname{span}(w_1^{\mathrm g},\ldots,w_{k-1}^{\mathrm g})
\right\}.
\end{equation}
$w_1^{\mathrm g}$ is just the Pauli vector attached to the largest Pauli coefficient of the input state, and the subsequent vectors are then 
\begin{equation}
w_k^{\mathrm g}
\in
\operatorname*{arg\,max}_{w\in\mathcal F_k}|r_w|,
\qquad
k=2,\ldots,m.
\end{equation}
Equivalently, we may sort all nonidentity Pauli coefficients once as $|r_{u_1}|\ge |r_{u_2}|\ge\cdots$ and, at each step, scan down this list until we find the first vector $u_\ell$ satisfying $L_{k-1}u_\ell=b_k$ and $
\operatorname{rank}
\begin{pmatrix}
w_1^{\mathrm g}&\cdots&w_{k-1}^{\mathrm g}&u_\ell
\end{pmatrix}
=k.$ The value obtained from this process is
\begin{equation}
M_{\mathrm g}
=
\sum_{k=1}^{m}|a_k|\,|r_{w_k^{\mathrm g}}|.
\end{equation}
The constructed tuple is linearly independent and has the same commutation matrix as $\mathbf v$. By theorem \ref{thm:pauli_set_orbit}, there exists a Clifford transformation mapping $v_k$ to $w_k^{\mathrm g}$ for every $k$. Thus $M_{\mathrm g}\le M_{\Omega}$: another admissible ordered tuple may attain a larger weighted sum. 

As seen in \cite{maity2026clifford}, we may also obtain an upper bound by ignoring the commutation constraints. Let $R_{(1)}\ge R_{(2)}\ge\cdots$ denote the sorted magnitudes of all non-identity Pauli coefficients of $\rho$. The rearrangement inequality gives
\begin{equation}
M_{\Omega}(\rho,\mathbf a)\le M_{\mathrm s}:=\sum_{k=1}^{m}|a_k|R_{(k)}.
\end{equation}
Combining the feasible value with the unconstrained sorting bound gives
\begin{equation}
\begin{aligned}
&\tr{\rho H}-a_0+M_{\mathrm g}\le\mathcal E_{\mathrm{Cl}}(\rho,H)\\&\mathcal E_{\mathrm{Cl}}(\rho,H)\le\tr{\rho H}-a_0+\sum_{k=1}^{m}|a_k|R_{(k)}.\end{aligned}
\end{equation}
The approximate value of the Clifford ergotropy is then 
\begin{equation}
   \mathcal E_{\mathrm{Cl}}^{\mathrm{est}}(\rho,H)=\mtr{\rho H}-a_0+\frac{M_{\mathrm s}+M_{\mathrm g}}{2}, 
\end{equation}
with absolute error at most $(M_{\mathrm s}-M_{\mathrm g})/2$. If the construction attains the weighted sorting bound, $M_{\mathrm g}=M_{\mathrm s}$, the Clifford ergotropy is obtained exactly. Finally, let $N=4^n-1$ be the number of nonidentity Pauli coefficients in the input state. Sorting these coefficients costs $\mathcal O(N\log N)$. A direct implementation that scans the sorted list at each of the $m$ stages and performs polynomial-time binary linear-algebra checks has worst-case complexity
\begin{equation}
\mathcal O\!\left(mN\,\operatorname{poly}(m,n)+N\log N\right).
\end{equation}
For fixed $m$, this is polynomial in the size $N$ of the Pauli expansion, apart from the initial sorting factor. If the coefficients are already sorted, the coefficient-processing cost is linear in $N$ for fixed $m$. Given the Pauli coefficients, this avoids exhaustive tuple enumeration; its dependence on $m$ is polynomial, while its dependence on $n$ remains exponential through $N=4^n-1$.

\section{Conclusion}
\label{sec:conclusion}

In this work, we have considered the problem of calculating the Clifford ergotropy. Since the Clifford group grows super-exponentially with the number of qubits, naively calculating this quantity is computationally challenging. Solving the Clifford ergotropy analytically for a general Hamiltonian is equally infeasible. The nontrivial action of Clifford conjugation makes the general orbit difficult to classify. Moreover, even for a single qubit, the analytical expression for the Clifford ergotropy is operationally more cumbersome than brute-force evaluation of the Clifford ergotropy over the one-qubit Clifford group. This demonstrates that obtaining a closed-form expression does not necessarily produce the most efficient method of calculation. However, by exploiting the commutation and linear-dependence structures of several relevant classes of Hamiltonians, we significantly reduced the corresponding optimization.

In particular, using the binary symplectic formalism for the Clifford group, we showed that Hamiltonians composed of linearly independent Pauli elements admit an exact orbit classification that substantially reduces the search space required to calculate the Clifford ergotropy. Although exact optimization remains exponential in the number of Hamiltonian terms, we introduced an approximation scheme polynomial in the supplied Pauli-expansion size, which itself grows exponentially with the number of qubits. For Hamiltonians with additional commutation structure, such as mutually commuting or pairwise anticommuting Pauli terms, the Clifford orbit simplifies further and admits more efficient optimization procedures.

 Although we have shown that imposing commutation structure reduces the searched set, it is still generically exponential. It is of interest to develop optimization algorithms for the Clifford ergotropy. As stated in Sec.~\ref{subsec:calc_ergo}, the commutation relations naturally lend themselves to a graphical representation, offering a start towards optimization. Furthermore, future work could extend the orbit classification to Hamiltonians containing linearly dependent Pauli terms. Such Hamiltonians include sums over complete stabilizer groups, for which multiplicative relations among the Pauli terms constrain the signs and permutations appearing in the Clifford orbit. Finally, because evaluating the Clifford ergotropy generally requires a large collection of Pauli expectation values, future work could investigate whether classical shadows, low-rank or other structural assumptions, or adaptive measurement protocols permit experimental realization without requiring full state tomography~\cite{Huang2020,Gross2010,Huszar2012}.

The most natural application lies in stabilizer quantum error correction. Quantum error correction may be viewed as a Maxwell-demon-like process in which syndrome information is acquired and used to restore an encoded state \cite{Nielsen1998}. Adapting the Clifford ergotropy to this setting could quantify the energetic performance of recovery operations and reveal possible tradeoffs between logical fidelity and extractable work. Together, these directions suggest that the Clifford ergotropy provides not only a measure of restricted work extraction, but also a framework for studying the energetic role of stabilizer structure and magic across quantum technologies.

\acknowledgements{E.D acknowledges U.S. NSF under Grant No. OSI-2328774. S.D acknowledges support from the John Templeton Foundation under Grant No. 63626. This work was supported by the U.S. Department of Energy, Office of Basic Energy Sciences, Quantum Information Science program in Chemical Sciences, Geosciences, and Biosciences, under Award No. DE-SC0025997.}

\appendix

\section{Counting Independent Commuting Pauli Tuples}
\label{app:stabilizer_count}

We derive the number of ordered, linearly independent, mutually commuting $m$-tuples of unsigned $n$-qubit Pauli operators using the binary symplectic formalism. Suppose that $k$ linearly independent, mutually commuting Pauli vectors $v_1,\ldots,v_k$ have already been selected. The first step is to count the number of allowed vectors $v_{k+1}$ that commute with the previously selected vectors. Since the span $W_k=\operatorname{span}\{v_1,\ldots,v_k\}$ forms a $k$-dimensional totally isotropic subspace, the next vector $v_{k+1}$ must belong to the symplectic orthogonal complement $W_k^\perp$; otherwise, commutativity is not preserved. Since the symplectic form is nondegenerate, the dimension of the complement may be found by rank-nullity, yielding $\dim W_k^\perp=2n-k$. Since two elements exist per binary dimension, $|W_k^\perp|=2^{2n-k}$.

To additionally preserve linear independence, we must exclude the $2^k$ vectors belonging to $W_k$. Therefore, the number of admissible choices for $v_{k+1}$ is $2^{2n-k}-2^k$. Multiplying over the $m$ successive choices gives the expected equation
\begin{equation}
    \left|\operatorname{Stab}_m^+(\mathcal{P}_n)\right|
    =
    \prod_{k=0}^{m-1}
    \left(2^{2n-k}-2^k\right),
\end{equation}
where $0\leq m\leq n$.

\section{Proofs for independent Pauli Hamiltonians}
\label{app:independent_proofs}

\subsection{Symplectic equivalence of independent tuples}
\label{app:tuple_orbit_proof}
\begin{proof}[Proof of Theorem~\ref{thm:pauli_set_orbit}]
We start with the forward direction. Let there exist an $S \in\operatorname{Sp}(2n,2)$ such that $Sv_i=w_i$ for all $i=1,\ldots,m$. Then $\omega(v_i,v_j)=\omega(w_i,w_j)$ for all $i,j=1,\ldots,m$, since a symplectic transformation preserves the symplectic inner product.

For the reverse direction, let $\omega(v_i,v_j)=\omega(w_i,w_j)$ for all $i,j=1,\ldots,m$. Define the subspaces of $\mathbb{F}^{2n}_2$ by $V=\operatorname{span}(v_1,\ldots,v_m)$ and $W=\operatorname{span}(w_1,\ldots,w_m)$. Note that the sets, by construction, form bases of their respective subspaces. Then we define the transformation $T:V\rightarrow W$ on the basis by $Tv_i=w_i$ for $i=1,\ldots,m$. Let $x=\sum_i b_iv_i$ and $y=\sum_i c_iv_i$ be arbitrary elements of $V$. Then

\begin{equation}
\begin{split}
    \omega(Tx,Ty)&=\sum_{i,j}b_ic_j\omega(w_i,w_j)\\
    &=\sum_{i,j}b_ic_j\omega(v_i,v_j)
    =\omega(x,y)
\end{split}
\end{equation}
implying $T$ is an isometry between $V$ and $W$. Then, by Witt's lemma for symplectic spaces~\cite{Aschbacher2000}, we may extend $T$ to a symplectic transformation $S:\mathbb{F}^{2n}_2\rightarrow\mathbb{F}^{2n}_2$. Therefore, $S\in\operatorname{Sp}(2n,2)$ and, since $S|_{V}=T$, $Sv_i=Tv_i=w_i$ for all $i=1,\ldots,m$. This completes the reverse direction and the proof.
\end{proof}

\subsection{Independent sign freedom}
\label{app:sign_freedom_proof}
\begin{proof}[Proof of Theorem~\ref{thm:independent_sign_freedom}]
Write each desired sign as $ s_i=(-1)^{b_i} $ for $ b_i\in\mathbb{F}_2, $ and let $  b=(b_1,\ldots,b_m)^T.$ Conjugation by a Pauli operator satisfies
\begin{equation}
    \sigma(u)\sigma(w_i)\sigma(u)^\dagger
    =
    (-1)^{\omega(u,w_i)}\sigma(w_i),
\end{equation}
so the proof is completed if there exists a $u\in\mathbb{F}_2^{2n}$ satisfying
\begin{equation}
    \omega(u,w_i)=b_i
    \qquad
    \text{for every }i=1,\ldots,m.
    \label{eq:wanted_signs}
\end{equation}

Let $W$ be the $m\times 2n$ matrix with $i$th row $w_i^T$. Using the symplectic form $\omega(w_i,u)=w_i^TJu$ the system in Eq.~\eqref{eq:wanted_signs} may be written as
\begin{equation}
    WJu=b.
    \label{eq:sign_equation}
\end{equation}
Because $w_1,\ldots,w_m$ are linearly independent, $\operatorname{rank}(W)=m$, and since $J$ is an invertible matrix $\operatorname{rank}(WJ) =\operatorname{rank}(W)= m.$. 
The linear map $L:\mathbb{F}^{2n}_2 \mapsto \mathbb{F}^{m}_2$ defined by the matrix $u\longmapsto WJu$ is therefore surjective. Therefore for every $b\in\mathbb{F}_2^m$, there exists a solution $u$ to Eq.~\eqref{eq:sign_equation}. For this choice of $u$,
\begin{equation}
    \sigma(u)\sigma(w_i)\sigma(u)^\dagger
    =
    (-1)^{b_i}\sigma(w_i)
    =
    s_i\sigma(w_i)
\end{equation}
for every $i$, proving that all $2^m$ sign assignments are accessible.
\end{proof}

\subsection{Clifford orbit of an independent Hamiltonian}
\label{app:hamiltonian_orbit_proof}
\begin{proof}[Proof of Corollary~\ref{cor:independent_hamiltonian_orbit}]
Let $\mathcal{A}$ denote the set on the right-hand side of Eq.~\eqref{eq:independent_hamiltonian_clifford_orbit}. We prove the equality by showing both inclusions.

First, let $H'=CHC^\dagger \in\mathcal{O}_{\mathrm{Cl}}(H) $ for some $C\in\mathrm{Cl}_n$. Clifford conjugation maps Pauli operators to signed Pauli operators, so there exist signs $\eta_i\in\{\pm1\}$ and vectors $w_i\in\mathbb{F}_2^{2n}$ such that $C\sigma(v_i)C^\dagger=\eta_i\sigma(w_i).$ Therefore 
\begin{equation}
    H'=a_0I+\sum_{i=1}^{m}\eta_i a_i\sigma(w_i)
\end{equation}
Clifford conjugation preserves both
linear independence and the commutation matrix, so $\Omega(\vec{w})=\Omega(\vec{v})$. Thus, $H'\in\mathcal{A}$, meaning $ \mathcal{O}_{\mathrm{Cl}}(H)\subseteq\mathcal{A}.$ For the reverse inclusion, let
\begin{equation}
    H' =a_0I+ \sum_{i=1}^{m}s_i a_i\sigma(w_i)  \in\mathcal{A}.
\end{equation}
By definition of $\mathcal{A}$, the tuple $\vec{w}=(w_1,\ldots,w_m)$ is linearly independent and satisfies $\Omega(\vec{w})=\Omega(\vec{v}).$ By Theorem~\ref{thm:pauli_set_orbit}, there exists a Clifford unitary
$C_0$ such that
\begin{equation}
    C_0\sigma_i(v_i)C_0^\dagger =\eta_i\sigma(w_i),
    \qquad
    \eta_i\in\{\pm1\},
\end{equation}
for every $i=1,\ldots,m$. Consequently,
\begin{equation}
C_0HC_0^\dagger=a_0I+\sum_{i=1}^{m}\eta_i a_i\sigma(w_i).
\end{equation}
The signs $\eta_i$ produced by the $C_0$ transformation may not match the desired signs $s_i$. However, because the vectors $w_1,\ldots,w_m$ are linearly independent, theorem~\ref{thm:independent_sign_freedom} guarantees the
existence of a Pauli operator $\sigma(u)$ satisfying
\begin{equation}
    \sigma(u)\sigma(w_i)\sigma(u)^\dagger = s_i\eta_i\sigma(w_i)
\end{equation}
for every $i$. Since $\eta_i^2=1$, conjugating the image of $\sigma(v_i)$
gives
\begin{align}
    \sigma(u)
    \left[
        \eta_i\sigma(w_i)
    \right]
    \sigma(u)^\dagger
    &=
    \eta_i(s_i\eta_i)\sigma(w_i) =
    s_i\sigma(w_i).
\end{align}
Therefore, the Clifford unitary $C=\sigma(u)C_0 $ maps $H \mapsto H'$, so $H'\in\mathcal{O}_{\mathrm{Cl}}(H)$, and consequently $\mathcal{A}\subseteq\mathcal{O}_{\mathrm{Cl}}(H). $ Combining the two inclusions gives $\mathcal{O}_{\mathrm{Cl}}(H)=\mathcal{A},$ which completes the proof.
\end{proof}

\bibliography{apssamp}

@book{NielsenChuang2010,
  author    = {Michael A. Nielsen and Isaac L. Chuang},
  title     = {Quantum Computation and Quantum Information},
  edition   = {10th Anniversary Edition},
  publisher = {Cambridge University Press},
  address   = {Cambridge},
  year      = {2010},
  isbn       = {9781107002173}
}

@Article{Sone2021entropy,
AUTHOR = {Sone, Akira and Deffner, Sebastian},
TITLE = {Quantum and Classical Ergotropy from Relative Entropies},
JOURNAL = {Entropy},
VOLUME = {23},
YEAR = {2021},
NUMBER = {9},
ARTICLE-NUMBER = {1107},
URL = {https://www.mdpi.com/1099-4300/23/9/1107},
PubMedID = {34573732},
ISSN = {1099-4300},
DOI = {10.3390/e23091107}
}

@article{Touil2022JPA,
doi = {10.1088/1751-8121/ac3eba},
url = {https://doi.org/10.1088/1751-8121/ac3eba},
year = {2021},
month = {dec},
publisher = {IOP Publishing},
volume = {55},
number = {2},
pages = {025301},
author = {Touil, Akram and Çakmak, Barış and Deffner, Sebastian},
title = {Ergotropy from quantum and classical correlations},
journal = {J. Phys. A: Math. Theor.}
}

@article{Kalsi2026fqxi,
  author  = {Kalsi, Tara and Sone, Akira and Bonan\c{c}a, M. V. S and Deffner, Sebastian},
  title   = {Thermodynamic work in the quantum regime},
  journal = {arXiv},
  note = {\textit{to be submitted}}
}

@article{Llobet2015PRX,
  title = {Extractable Work from Correlations},
  author = {Perarnau-Llobet, Mart{\'{\i}} and Hovhannisyan, Karen V. and Huber, Marcus and Skrzypczyk, Paul and Brunner, Nicolas and Ac{\'{\i}}n, Antonio},
  journal = {Phys. Rev. X},
  volume = {5},
  issue = {4},
  pages = {041011},
  numpages = {14},
  year = {2015},
  month = {Oct},
  publisher = {American Physical Society},
  doi = {10.1103/PhysRevX.5.041011},
  url = {https://link.aps.org/doi/10.1103/PhysRevX.5.041011}
}

@article{Myers2021PRXQ,
  title = {Thermodynamics of Statistical Anyons},
  author = {Myers, Nathan M. and Deffner, Sebastian},
  journal = {PRX Quantum},
  volume = {2},
  issue = {4},
  pages = {040312},
  numpages = {28},
  year = {2021},
  month = {Oct},
  publisher = {American Physical Society},
  doi = {10.1103/PRXQuantum.2.040312},
  url = {https://link.aps.org/doi/10.1103/PRXQuantum.2.040312}
}

@Article{Touil2021Entropy,
AUTHOR = {Touil, Akram and Weber, Kevin and Deffner, Sebastian},
TITLE = {Quantum Euler Relation for Local Measurements},
JOURNAL = {Entropy},
VOLUME = {23},
YEAR = {2021},
NUMBER = {7},
ARTICLE-NUMBER = {889},
URL = {https://www.mdpi.com/1099-4300/23/7/889},
PubMedID = {34356429},
ISSN = {1099-4300},
DOI = {10.3390/e23070889}
}

@article{Alipour2016SR,
	author = {Alipour, S. and Benatti, F. and Bakhshinezhad, F. and Afsary, M. and Marcantoni, S. and Rezakhani, A. T.},
	date = {2016/10/21},
	doi = {10.1038/srep35568},
	id = {Alipour2016},
	isbn = {2045-2322},
	journal = {Sci. Rep.},
	number = {1},
	pages = {35568},
	title = {Correlations in quantum thermodynamics: Heat, work, and entropy production},
	url = {https://doi.org/10.1038/srep35568},
	volume = {6},
	year = {2016}}

@article{Aguilar2025SA,
author = {Milton Aguilar  and Eric Lutz },
title = {Correlated quantum machines beyond the standard second law},
journal = {Sci. Adv.},
volume = {11},
number = {41},
pages = {eadw8462},
year = {2025},
doi = {10.1126/sciadv.adw8462},
URL = {https://www.science.org/doi/abs/10.1126/sciadv.adw8462}}

@misc{maity2026clifford,
      title={Clifford Ergotropy}, 
      author={Somnath Maity and Ryusuke Hamazaki},
      year={2026},
      eprint={2605.09878},
      archivePrefix={arXiv},
      primaryClass={quant-ph},
      url={https://arxiv.org/abs/2605.09878}
}

@phdthesis{Gottesman1997,
    author  = {Gottesman, Daniel},
    title   = {Stabilizer Codes and Quantum Error Correction},
    school  = {California Institute of Technology},
    year    = {1997},
    eprint  = {quant-ph/9705052},
    archivePrefix = {arXiv},
    primaryClass  = {quant-ph}
}

@article{DehaeneDeMoor2003,
    author  = {Dehaene, Jeroen and De Moor, Bart},
    title   = {Clifford Group, Stabilizer States, and Linear and
               Quadratic Operations over {GF}(2)},
    journal = {Phys. Rev. A},
    volume  = {68},
    number  = {4},
    pages   = {042318},
    year    = {2003},
    doi     = {10.1103/PhysRevA.68.042318}
}

@article{ullmann1976algorithm,
  author  = {Ullmann, J. R.},
  title   = {An Algorithm for Subgraph Isomorphism},
  journal = {Journal of the ACM},
  volume  = {23},
  number  = {1},
  pages   = {31--42},
  year    = {1976},
  doi     = {10.1145/321921.321925}
}

@article{cordella2004subgraph,
  author  = {Cordella, Luigi P. and Foggia, Pasquale and
             Sansone, Carlo and Vento, Mario},
  title   = {A (Sub)Graph Isomorphism Algorithm for Matching
             Large Graphs},
  journal = {IEEE Transactions on Pattern Analysis and
             Machine Intelligence},
  volume  = {26},
  number  = {10},
  pages   = {1367--1372},
  year    = {2004},
  doi     = {10.1109/TPAMI.2004.75}
}

@book{schrijver2003combinatorial,
  author    = {Schrijver, Alexander},
  title     = {Combinatorial Optimization: Polyhedra and Efficiency},
  publisher = {Springer},
  address   = {Berlin},
  year      = {2003}
}

@article{kitaev2003fault,
  author  = {Kitaev, A. Yu.},
  title   = {Fault-Tolerant Quantum Computation by Anyons},
  journal = {Ann. Phys.},
  volume  = {303},
  number  = {1},
  pages   = {2--30},
  year    = {2003},
  doi     = {10.1016/S0003-4916(02)00018-0}
}

@article{raussendorf2001oneway,
  author  = {Raussendorf, Robert and Briegel, Hans J.},
  title   = {A One-Way Quantum Computer},
  journal = {Phys. Rev. Lett.},
  volume  = {86},
  number  = {22},
  pages   = {5188--5191},
  year    = {2001},
  doi     = {10.1103/PhysRevLett.86.5188}
}

@article{dirac1928quantum,
  author  = {Dirac, P. A. M.},
  title   = {The Quantum Theory of the Electron},
  journal = {Proc. R. Soc. A},
  volume  = {117},
  number  = {778},
  pages   = {610--624},
  year    = {1928},
  doi     = {10.1098/rspa.1928.0023}
}

@article{elliott2015majorana,
  author  = {Elliott, Steven R. and Franz, Marcel},
  title   = {Colloquium: Majorana Fermions in Nuclear,
             Particle, and Solid-State Physics},
  journal = {Rev. Mod. Phys.},
  volume  = {87},
  number  = {1},
  pages   = {137--163},
  year    = {2015},
  doi     = {10.1103/RevModPhys.87.137}
}

@article{izmaylov2020unitary,
  author  = {Izmaylov, Artur F. and Yen, Tzu-Ching and
             Lang, Robert A. and Verteletskyi, Vladyslav},
  title   = {Unitary Partitioning Approach to the Measurement
             Problem in the Variational Quantum Eigensolver Method},
  journal = {J. Chem. Theory Comput.},
  volume  = {16},
  number  = {1},
  pages   = {190--195},
  year    = {2020},
  doi     = {10.1021/acs.jctc.9b00791}
}

@book{Aschbacher2000,
  author    = {Michael Aschbacher},
  title     = {Finite Group Theory},
  edition   = {2},
  publisher = {Cambridge University Press},
  year      = {2000}
}

@article{Huang2020,
  author  = {Huang, Hsin-Yuan and Kueng, Richard and Preskill, John},
  title   = {Predicting Many Properties of a Quantum System from Very Few Measurements},
  journal = {Nature Physics},
  volume  = {16},
  pages   = {1050--1057},
  year    = {2020},
  doi     = {10.1038/s41567-020-0932-7}
}

@article{Gross2010,
  author  = {Gross, David and Liu, Yi-Kai and Flammia, Steven T. and Becker, Stephen and Eisert, Jens},
  title   = {Quantum State Tomography via Compressed Sensing},
  journal = {Phys. Rev. Lett.},
  volume  = {105},
  pages   = {150401},
  year    = {2010},
  doi     = {10.1103/PhysRevLett.105.150401}
}

@article{Huszar2012,
  author  = {Husz{\'a}r, Ferenc and Houlsby, N. M. T.},
  title   = {Adaptive Bayesian Quantum Tomography},
  journal = {Phys. Rev. A},
  volume  = {85},
  pages   = {052120},
  year    = {2012},
  doi     = {10.1103/PhysRevA.85.052120}
}

@article{Allahverdyan2004,
  author  = {Allahverdyan, A. E. and Balian, R. and Nieuwenhuizen, Th. M.},
  title   = {Maximal Work Extraction from Finite Quantum Systems},
  journal = {EPL (Europhys. Lett.)},
  volume  = {67},
  number  = {4},
  pages   = {565--571},
  year    = {2004},
  doi     = {10.1209/epl/i2004-10101-2}
}

@article{Lostaglio2015,
  author  = {Lostaglio, Matteo and Korzekwa, Kamil and Jennings, David and Rudolph, Terry},
  title   = {Quantum Coherence, Time-Translation Symmetry, and Thermodynamics},
  journal = {Phys. Rev. X},
  volume  = {5},
  number  = {2},
  pages   = {021001},
  year    = {2015},
  doi     = {10.1103/PhysRevX.5.021001}
}

@article{Francica2017,
  author  = {Francica, Gianluca and Goold, John and Plastina, Francesco and Paternostro, Mauro},
  title   = {Daemonic Ergotropy: Enhanced Work Extraction from Quantum Correlations},
  journal = {npj Quantum Inf.},
  volume  = {3},
  pages   = {12},
  year    = {2017},
  doi     = {10.1038/s41534-017-0012-8}
}

@article{Goold2016,
  author  = {Goold, John and Huber, Marcus and Riera, Arnau and Skrzypczyk, Paul and del Rio, Lidia},
  title   = {The Role of Quantum Information in Thermodynamics---A Topical Review},
  journal = {J. Phys. A: Math. Theor.},
  volume  = {49},
  number  = {14},
  pages   = {143001},
  year    = {2016},
  doi     = {10.1088/1751-8113/49/14/143001}
}

@article{Salvia2023,
  author  = {Salvia, Raffaele and De Palma, Giacomo and Giovannetti, Vittorio},
  title   = {Optimal Local Work Extraction from Bipartite Quantum Systems
             in the Presence of Hamiltonian Couplings},
  journal = {Phys. Rev. A},
  volume  = {107},
  number  = {1},
  pages   = {012405},
  year    = {2023},
  doi     = {10.1103/PhysRevA.107.012405}
}

@article{Brown2016,
  author  = {Brown, Eric G. and Friis, Nicolai and Huber, Marcus},
  title   = {Passivity and Practical Work Extraction Using Gaussian Operations},
  journal = {New J. Phys.},
  volume  = {18},
  number  = {11},
  pages   = {113028},
  year    = {2016},
  doi     = {10.1088/1367-2630/18/11/113028}
}

@article{Castellano2024,
  author  = {Castellano, Riccardo and Farina, Donato and Giovannetti, Vittorio
             and Ac{\'i}n, Antonio},
  title   = {Extended Local Ergotropy},
  journal = {Phys. Rev. Lett.},
  volume  = {133},
  number  = {15},
  pages   = {150402},
  year    = {2024},
  doi     = {10.1103/PhysRevLett.133.150402}
}

@article{Beverland2021,
  author  = {Beverland, Michael E. and Kubica, Aleksander and Svore, Krysta M.},
  title   = {Cost of Universality: A Comparative Study of the Overhead of State Distillation and Code Switching with Color Codes},
  journal = {PRX Quantum},
  volume  = {2},
  number  = {2},
  pages   = {020341},
  year    = {2021},
  doi     = {10.1103/PRXQuantum.2.020341}
}

@article{Aaronson2004,
  author  = {Aaronson, Scott and Gottesman, Daniel},
  title   = {Improved Simulation of Stabilizer Circuits},
  journal = {Phys. Rev. A},
  volume  = {70},
  number  = {5},
  pages   = {052328},
  year    = {2004},
  doi     = {10.1103/PhysRevA.70.052328}
}

@article{Bravyi2005,
  author  = {Bravyi, Sergey and Kitaev, Alexei},
  title   = {Universal Quantum Computation with Ideal Clifford Gates and Noisy Ancillas},
  journal = {Phys. Rev. A},
  volume  = {71},
  number  = {2},
  pages   = {022316},
  year    = {2005},
  doi     = {10.1103/PhysRevA.71.022316}
}

@article{Koenig2014,
  author  = {Koenig, Robert and Smolin, John A.},
  title   = {How to efficiently select an arbitrary Clifford group element},
  journal = {J. Math. Phys.},
  volume  = {55},
  number  = {12},
  pages   = {122202},
  year    = {2014},
  doi     = {10.1063/1.4903507}
}

@article{Nielsen1998,
  author  = {Nielsen, Michael A. and Caves, Carlton M.
             and Schumacher, Benjamin and Barnum, Howard},
  title   = {Information-Theoretic Approach to Quantum Error
             Correction and Reversible Measurement},
  journal = {Proc. R. Soc. Lond. A},
  volume  = {454},
  number  = {1969},
  pages   = {277--304},
  year    = {1998},
  doi     = {10.1098/rspa.1998.0160}
}

\end{document}